\documentstyle[preprint,aps]{revtex}
 \begin{document}
 \begin{title}
{\bf Energy spectrum, density of states and optical transitions \\
in strongly biased narrow-gap quantum wells}
\end{title}
\author{ O. E. Raichev\cite{oleg}, P. Vasilopoulos\cite{takis}, and
F.T.Vasko$^{\ddagger}$ }
\address{\cite{takis}Concordia University, Department of Physics,\\ 1455 de
Maisonneuve Blvd. Ouest,  Montr\'{e}al, Qu\'{e}bec, Canada, H3G 1M8
\ \\
\ \\
\cite{oleg} $^{\ddagger}$Institute of Semiconductor Physics, National Academy
of Sciences of Ukraine,\\  Pr. Nauki 45, Kiev-28, 252650, Ukraine}
\maketitle
\begin{abstract}

We study theoretically the effect of an electric field on the electron
states and far-infrared optical properties in narrow-gap lead salt quantum
wells. The electron states are described by a two-band Hamiltonian.
An application of a strong electric field across the
well allows the control of the energy gap between the two-dimensional (2D)
states in a wide range. A sufficiently strong electric field transforms the
narrow-gap quantum well to a nearly gapless 2D system, whose electron energy
spectrum is described by linear dispersion relations $\varepsilon_{\sigma}
(k) \sim \pm (k-k_{\sigma})$, where $k_{\sigma}$ are the field-dependent 2D
momenta corresponding to the minimum energy gaps for the states with
spin numbers $\sigma$. Due to the field-induced shift of the 2D subband
extrema away from $k=0$ the density of states has inverse-square-root
divergencies at the edges. This property may result in a considerable
increase of the magnitude of the optical absorption and in the efficiency of
the electrooptical effect.\\


\end{abstract}


\section{Introduction}

The modification of electron spectra and of the optical absorption of
semiconductor heterostructures by strong electric fields applied parallel
to the growth direction (see band diagram in Fig.1) has important applications
in infrared optical modulators and tunable photodetectors. Although the
basic physics of these phenomena is related to the quantum confined Stark
effect, their rigorous examination should take into account the complex band
structure of the materials under investigation, especially in the case
of interband optical transitions. The electric-field effect on electron
spectra and optical absorption in quantum wells has been studied for a
wide class of the materials: AlInAs/GaInAs$^{1}$, Si/Ge$^{2}$, and
narrow-gap CdTe/HgTe$^{3}$ and InAs/AlGaSb$^4$
systems. The use of narrow-gap materials is especially important in
connection with an increasing interest in the optics of mid- and
far-infrared regions. For this reason, heterostructures based on
narrow-gap lead salt alloys$^{5}$ may become promising.

The electron spectrum of the lead salt alloys at the L-points
of the Brillouin zone is described by an anisotropic two-band Dirac-like
Hamiltonian and the energy gap strongly depends on the alloy composition.
The most extensively investigated alloy, Pb$_{1-x}$Sn$_x$Te, becomes a
gapless material at $x \simeq 0.47$. Electronic properties including
energy spectrum, optical transitions and electron transport of multilayer
heterostructures based upon Pb$_{1-x}$Sn$_x$Te materials, have been
studied in Refs. 6-8. Theoretical investigations of the electron energy
spectrum of a single PbSnTe quantum well are summarized in Ref. 9, and
interband optical transitions in such quantum wells have been studied
in Ref. 10. The effect of weak electric fields on the energy spectrum
and optical absorption has been studied in Ref. 11, where approximate
analytical solutions of the eigenvalue problem have been found by
perturbation theory. In that case, the Stark shifts of the levels
are quadratic in the electric field and comparable to the spin
splitting energies due to field-induced asymmetry of the confining
potential.

In this paper we study theoretically  the influence of a strong
electric field on the energy spectrum, density of states and optical
absorption of quantum wells made from narrow-gap lead salt alloys. Our
investigation is motivated as follows. From the point of view of fundamental
physics it is important to study the eigenvalue problem for the quantum well
in the strong-field regime, when the drop of the electrostatic potential
energy across the well becomes comparable to both the size-quantized energies
and the energy gap of the quantum-well material. As we show below, the
electron spectrum in such conditions has   a
pronounced spin splitting and an anticrossing of two-dimensional (2D)
conduction- and valence-band states as they are shifted towards each
other by the electric field. As a result, a nearly gapless 2D electron
system is formed. The spectra of both spin-up and spin-down states become
non-monotonic, the positions of the 2D subband extrema in the momentum space
are moved by the electric field so that the density of states acquires
inverse-square-root divergencies near its edges. These properties are also
important from the point of view of applied physics because an application
of a strong electric field allows one to vary the energy gap between
conduction and valence subbands in a wide range, sweeping the energy of
fundamental optical transitions across the entire far-infrared region.
Moreover, due to the described peculiarity of the density of states,
the optical absorption is strongly enhanced in the frequency regions
corresponding to the optical transitions of electrons near the subband
extrema.

The paper is organized as follows. In Sec. II we present the basic
formalism for the calculation of the spectrum, the wave functions, and the
density of states in the narrow-gap quantum wells. In Sec. III we show the
results of a numerical calculation of the electron spectrum and of the
density of states associated with the ground-state conduction and valence
subbands, discuss their properties in strong fields, and support our
discussion by analytical considerations. In Sec. IV we calculate the
infrared optical absorption coefficient and its field dependence
(electrooptic effect) for different polarizations of the incident
radiation and discuss the characteristic features of the absorption.
In Sec. V we make concluding remarks. The Appendix contains general
expressions for the wave functions of the two-band model in terms
of standard special functions.

\section{General formalism}

Within the ${\bf k} \cdot {\bf p}$ -formalism, the electron
states near the L-points of the Brillouin zone in lead salt alloys are
described by an anisotropic two-band Hamiltonian. Below we assume
that the heterostructure containing a quantum well is grown along the
(111) crystallographic direction and consider the states belonging to
those two valleys whose large-effective-mass axis is (111). The states
belonging to the other six valleys are characterized by a considerably
smaller effective mass in the growth direction and, as a result, have
higher size-quantization energies$^{12}$. Since we are interested in
the ground valence- and conduction-band states, we remove these
high-energy states from consideration. Thus, the columnar eigenstates
$\Psi =(\varphi, \chi)$ in the region inside the quantum well ($-d/2
< z < d/2$) satisfy the matrix equation $(\hat{H}-\varepsilon) \Psi=0$
written explicitly as (in all equations or formulas we put
$e=\hbar=1$)

\begin{equation}
\left( \begin{array}{cc}
\frac{\Delta}{2} + F z -\varepsilon & P_{||}
\frac{\partial}{\partial z} + P_{\bot} \sigma k \\
-P_{||} \frac{\partial}{\partial z} + P_{\bot} \sigma k &
-\frac{\Delta}{2} + F z -\varepsilon
\end{array} \right)
\left( \begin{array}{c}
\varphi (z) \\ \chi (z)
\end{array} \right)=0~.
\label{1}
\end{equation}
Here $\Delta$ is the energy gap of the material inside the
quantum well, $F$ is the magnitude of the applied electric
field, $\sigma=\pm 1$ is the spin quantum number, $k$ is the
absolute value of 2D momentum (here and below the 2D momentum
is counted 
from the L-point), and $P_{||}$ and $P_{\bot}$
are the matrix elements of velocity operators in the directions
parallel and perpendicular to (111) axis. The energy $\varepsilon$
is counted from the middle point of the band gap. A general solution
of Eq. (1) can be expressed in terms of the special functions, see
Appendix for details.

Equation (1) represents a set of two first-order differential equations.
The most general boundary conditions for such a set have the form
\begin{equation}
\chi (-d/2)= \alpha_l \varphi (-d/2),~~\chi (d/2)= \alpha_r
\varphi (d/2)~,
\end{equation}
where the coeffecients $\alpha_l$ and $\alpha_r$ characterize the left
and right boundaries. If the barriers of the quantum-well
heterostructure are made from a material which is also described
by a two-band Hamiltonian, cf. Fig. 1, an explicit form of these
coefficients can be derived from the requirement of continuity of
both $\varphi (z)$ and $\chi (z)$ across the boundaries. A simple
calculation, done under the assumption that $P_{||}$ and $P_{\bot}$
do not change across the interfaces, gives

\begin{eqnarray}
\alpha_r= \sqrt{u_c^+/u_v^- + (P_{\bot}k/u_v^-)^2 } + P_{\bot} \sigma k/u_v^-
\\
\alpha_l=-\sqrt{u_c^-/u_v^+ + (P_{\bot}k/u_v^+)^2 } + P_{\bot} \sigma k/u_v^+
\end{eqnarray}
where $u_c^{\pm}=U_c +\Delta/2 \pm Fd/2 -\varepsilon$ and $u_v^{\pm}=
U_v +\Delta/2 \pm Fd/2 +\varepsilon$, and $U_c$ and $U_v$ are the conduction-
and valence-band offsets defined in Fig. 1. If the barriers are high enough
to neglect $\Delta$, $Fd$, $\varepsilon$ and $P_{\bot} k$ in comparison to
$U_c$ and $U_v$, Eqs. (3) and (4) take a simple form$^{11}$ $\alpha_{r,l}= \pm
\sqrt{U_c/U_v}$.

If $F=0$, Eq. (1) with boundary conditions (2) is solved exactly,
see also Ref. 9. The wave functions take the form $\varphi= c_1
e^{i q z} + c_2 e^{-i q z}$ and $\chi= c_3 e^{i q z} + c_4
e^{-i q z}$, where $q =\sqrt{\varepsilon^2-\Delta^2/4
- P_{\bot}^2 k^2 }/P_{||} $. An application of the boundary
conditions leads to the dispersion relation
$$
\tan(q d) = \frac{(\alpha_r-\alpha_l) P_{||} q }{(1+\alpha_l \alpha_r)
\varepsilon-(1-\alpha_l \alpha_r) \Delta/2 - (\alpha_l+\alpha_r)
P_{\bot} \sigma k}
$$
describing a set of nonparabolic 2D subbands. Even if the quantum well
is formed by a gapless semiconductor ($\Delta=0$), the size quantization
effect always produces a finite energy gap between the valence- and
conduction-band states.

Equation (1) has an important symmetry property. It is
invariant with respect to the transformations $\varepsilon
\rightarrow - \varepsilon$, $z \rightarrow -z$, $\varphi (z)
\rightarrow \chi (-z)$, and $\chi (z) \rightarrow -\varphi
(-z)$. This means that the energy spectrum of electrons
is symmetric with respect to $\varepsilon =0$ if the
boundary conditions allow the above-described transformation
of the wave function. Therefore, the symmetry of the spectrum exists for

\begin{equation}
\alpha_l (\varepsilon) \alpha_r (-\varepsilon) = -1 .
\end{equation}
If the boundary coefficients are given by Eqs. (3) and
(4), Eq. (5) is equivalent to $U_c=U_v$. The symmetry of
the spectra in such conditions is evident from a qualitative
point of view.

Another important property is the removal of the spin degeneracy
by the electric field $F$. If $\alpha_l (\varepsilon)= -
\alpha_r (\varepsilon)$, i.e., if the left and right boundary
potentials are equal,
and $F=0$, the spectra of the states with different spin numbers
are the same. Applying an   electric field introduces
an asymmetric potential and renders the spectra of $\sigma=1$ and
$\sigma=-1$ states different. This removal of spin degeneracy is
a consequence of the spin-orbit interaction and it is typical
for asymmetric  heterostructures$^{13}$.

If the electron energy spectrum $\varepsilon_{n \sigma}(k)$ of
the Hamiltonian of Eq. (1)  is known, one can find the density of electron
states
$\rho(\varepsilon)$ according to

\begin{equation}
\rho(\varepsilon)=\sum_{n \sigma} \int_0^{\infty} \frac{k dk}{2 \pi}
\delta (\varepsilon - \varepsilon_{n \sigma}(k) ).
\end{equation}
Using the well-known identity $\partial \varepsilon_{n \sigma} (k)/
\partial k = \left< \Psi_{n \sigma k} | \partial\hat{H}/\partial k
| \Psi_{n \sigma k} \right>$, and taking   advantage of
$\partial\hat{H}/\partial k = \sigma P_{\bot} \hat{\sigma}_x$, where
$\hat{\sigma}_x$ is a Pauli matrix, we can rewrite Eq. (6) as

\begin{equation}
\rho(\varepsilon)=\sum_{n \sigma i}
\frac{k_i}{4 \pi P_{\bot}} \left| \int d z \varphi_{n \sigma k_i}(z)
\chi_{n \sigma k_i}(z) \right|^{-1},
\end{equation}
where $k_i$ are the roots of the equation $\varepsilon =
\varepsilon_{n \sigma}(k)$.

\section{Energy spectrum and density of states}

As shown in the Appendix the general solution of Eq. (1) can be
expressed in terms of the parabolic cylinder functions. However,
the analysis of the results becomes more difficult than a direct
numerical solution of Eqs.  (1) and (2). Below we describe the
results of such a direct solution based on the application of the 4th order
Runge-Kutta procedure to Eq. (1) and subsequent determination of
the energies which satisfy Eq. (2). In all calculations we use
interband velocities taken from Ref. 6: $\hbar P_{||}=0.14$ eV nm and
$\hbar P_{\bot} = 0.47$ eV nm.
Figures 2 and 3 show the evolution of the spectrum and
of the density of states for two ground-state 2D subbands with the
increase of the electric field $F$ at $d=12$ nm, $\alpha_l=-1$, and
$\alpha_r=1$.  The chosen boundary conditions may correspond to a
symmetric heterostructure with infinitely large but equal conduction-
and valence-band offsets.  For the sake of convenience $\varepsilon_{n
\sigma}(k)$ is plotted as a function of $\sigma k$ so that the left
sides of Figs.  2 (a) and 3 (a) describe the states with $\sigma=-1$,
while the right sides describe those with $\sigma=1$.  Regardless of
the gap widths, the electric field introduces similar qualitative
changes to the spectrum.

First we see Stark shifts and spin splitting. The 2D subband extrema for
the $\sigma=-1$ states are shifted away from $k=0$. Further, this shift
becomes
more pronounced and it is accompanied with a substantial decrease of the
gap between the 2D subbands. When the field is stronger than 2 $\times$
10$^5$ V/cm, the gap becomes very small, typically 1-2 meV, and the subband
dispersion in the vicinity of the gap becomes linear. The position of the
energy extrema in  momentum space is given by a characteristic momentum
$k_{-}$, which increases almost linearly with the increase of the field.
This shift is accompanied with an exponentially fast decrease of the gap.
At higher fields, this kind of behavior repeats itself for the $\sigma=1$
states. Even at weak fields there exists a shift of the subband extrema
away from $k=0$. It produces an inverse-square-root divergence of the
density of states at the edge. At stronger fields, when the electron
dispersion for both spin-up and spin-down states becomes non-monotonic,
similar divergences also appear near the extrema points above the edges
so that the density of states becomes rather complex. Figures 2 (b) and 3
(b) also show bump-like features of the density of states for the energies
corresponding to zero momentum because the states with different spins
give different contributions to the density of states above and below
these energies. A similar calculation for narrower wells demonstrates
all the features explained above. However, they appear at higher fields
than for wider wells.

For an explanation of the small gaps and linear energy spectra near the extrema
points in strong electric fields, it is instructive to take a look at the
wave functions $(\varphi, \chi)$ for conduction and valence subbands at
$k=k_{-}$, see Fig. 4. The electron density is high near the interfaces
and low near the center of the well. In other words, the strong electric
field makes the electron states at $k \sim k_{-}$ localized in narrow regions
near the interfaces. An analysis of Eq. (1) leads to the same conclusion.
For more clarity we suppose that the condition (5) is fulfilled so it is
sufficient for our purposes to consider small-energy solutions, $\varepsilon
\ll Fd$. Let us consider the region near the left boundary and introduce
$z=-d/2 + d y$, where $y \ll 1$. Then, expressing $\chi$ through $\varphi$ we
reduce Eq. (1) to

\begin{equation}
-\frac{\partial^2}{\partial y^2} \varphi(y) + \left[
\left(\frac{Fd^2}{P_{||}}\right)^2 y - \lambda \right] \varphi(y)=0,
\end{equation}
where

\begin{equation}
\lambda = R(\sigma k,\varepsilon) \equiv \left( \frac{Fd^2}{2P_{||}}
\right)^2 -
\left( \frac{P_{\bot}dk}{P_{||}} \right)^2 -\left( \frac{\Delta d}{2P_{||}}
\right)^2 - 2 \frac{P_{\bot}d \sigma k}{P_{||}(1+\Delta/Fd)} +
\frac{Fd^3}{P_{||}^2} \varepsilon,
\end{equation}
while the boundary condition for $\varphi(y)$ at $y=0$ ($z=-d/2$)
becomes

\begin{equation}
\left[ \left. \frac{\partial }{\partial y} + \alpha_l \frac{Fd^2}{2 P_{||}}
\left(1+\frac{\Delta}{Fd} \right) - \frac{P_{\bot}d \sigma k}{P_{||}} \right]
\varphi(y) \right|_{y=0}=0.
\end{equation}
Equation (8) is valid for $y \ll 1$. To make a characteristic
spatial scale of its solutions small in comparison to unity,
we should also impose the requirement $(Fd^2/P_{||})^{2/3} \gg 1$,
which is fulfilled for $d \sim 10$ nm and $F > 10^5$ eV/cm.
We also assume that $Fd > \Delta$.

Equation (8) with the boundary condition (10) describes a one-dimensional
quantum-mechanical problem for the spectrum and eigenfunctions of a particle
localized in a triangular potential well formed by a barrier at $y=0$
and uniform field at $y > 0$. It has discrete solutions $\lambda=
\lambda_{n \sigma}(k)$, from which we take the ground-state solutions
$\lambda_{0 \sigma}(k)$ corresponding to the nodeless wave functions
$\varphi_{0 \sigma k}(y)$. If we solve the equation $\lambda_{0
\sigma}(k)= R(\sigma k,\varepsilon)$ for small energies, we obtain
a linear spectrum $\varepsilon_{\sigma}(k) = A_{\sigma}(k -k_{\sigma})$,
where the characteristic velocities $A_{\sigma}$ are of the order of
$P_{\bot}$ and the characteristic momenta $k_{\sigma}$, corresponding to
$\varepsilon_{\sigma}(k)=0$, are of the order of $Fd/P_{\bot}$. A similar
consideration can be done for the right boundary by introducing 
  $z=d/2-d y$. Due to the symmetry of the problem, the equations
for the right boundary are simply obtained  from Eqs. (8)-(10) by  replacing
 $\varphi $ 
 by $ \chi$ and $\varepsilon$ 
by $-\varepsilon$ only. The second replacement is significant, because
it leads to $\varepsilon_{\sigma}(k) = - A_{\sigma}(k -k_{\sigma})$.
Thus, for each spin number we have two linear branches intersecting
each other at $\varepsilon=0$ and $k = k_{\sigma}$. This description
would be completely correct if there were no overlap between the
states localized near each boundary. Really, such an overlap always
exists, and it produces an anticrossing near $\varepsilon=0$. As a
result, a small but finite energy gap appears. This gap becomes
progressively smaller with the increase of the electric field. From a
qualitative point of view, the described effect can be explained as
a result of the anticrossing of the conduction and valence subbands
as they are shifted towards each other by the electric field.

Figure 2 demonstrates another remarkable feature, namely that
the energy spectra of a zero-gap quantum well at $k=0$
are independent of the
electric field, see Fig. 2 (a). This property can be rigorously
proved because Eq. (1) is exactly solvable at $\Delta=0$
and $k=0$, regardless of the strength of the electric field. In these
conditions Eq. (1) is diagonalized by the unitary transformation
$\varphi = (\tilde{ \varphi}+ i \tilde{ \chi})/\sqrt{2}$ and
$\chi = (\tilde{ \chi}+ i \tilde{ \varphi})/\sqrt{2}$; we then
obtain the equations
\begin{eqnarray}
\left[i P_{||} \frac{\partial}{\partial z} + F z -
\varepsilon \right] \tilde{ \varphi}=0, \\
\nonumber
\ \\
\left[-i P_{||} \frac{\partial}{\partial z} + F z -
\varepsilon \right] \tilde{ \chi}=0,
\end{eqnarray}
whose solutions are straightforward. We have
\begin{equation}
\varphi= C_1 \cos \psi_z + C_2 \sin \psi_z,~
\chi= C_2 \cos \psi_z - C_1 \sin \psi_z~,
\end{equation}
where $\psi_z=(Fz^2/2-\varepsilon z)/P_{||}$. Applying
the boundary conditions gives the dispersion relation
\begin{equation}
\tan \left(\frac{\varepsilon d}{P_{||}} \right)= \frac{\alpha_r -
\alpha_l}{1+\alpha_l \alpha_r}.
\end{equation}
Thus, the quantization energies described by Eq. (14) are independent
of the electric field if the boundary coefficients are field-independent.
If $\alpha_l$ and $\alpha_r$ are energy-independent, Eq. (14) describes
a ladder of equally spaced levels. Further, if Eq. (5) is fulfilled,
the spectrum at $k=0$ is always field-independent. It is given by
$\varepsilon = (n+1/2)(\pi P_{||}/d)$, where $n$ is an integer.

The electric fields required for  the anticrossing are rather
strong and the change of the electrostatic potentials across the well are
comparable to the band offsets $U_c$ and $U_v$ of PbSnTe heterostructure.
To demonstrate that the effects still exist for realistic band offsets,
we have done a calculation of the spectrum and density of states for a
quantum well made of a gapless Pb$_{1-x}$Sn$_x$Te alloy layer enclosed
by SnTe layers. For such a system, we used the parameters
$U_c=0.25$ eV and $U_v=0.08$ eV, taken from Ref. 6, and employed the
boundary conditions (3) and (4). The calculated energy spectrum for a
12 nm wide quantum well is shown in Fig. 5. Since Eq. (5) is no longer
valid in this case, both the spectrum and the density of states are
not symmetric with respect to $\varepsilon=0$. For this reason, we plot
the density of states for both   conduction and   valence subbands. Due to
the dependence of the boundary conditions on the energy, momentum, and
electric field, discrete solutions do not always exist. For example,
at $F=1.5 \times$ 10$^{5}$ V/cm the region of their existence becomes
rather narrow. Nevertheless, the salient features shown in Figs. 2 and 3
are reproduced in Fig. 5.

The divergences of the density of states arising near the extrema
of the electron energy spectra can be studied by different means,
including measurements of the thermodynamic quantities such as specific
heat and capacitance, optical transmission and absorption  measurements, and
by transport measurements. Considering possible applications of the biased
narrow-gap quantum wells to optical detectors and modulators, in the next
section we evaluate the spectral and field dependence of the far-infrared
optical absorption coefficient and show how the special features of the
spectrum manifest themselves in the optical transitions.

\section{Far-infrared optical transitions}

In the dipole approximation, the optical absorption coefficient
$\xi$ is given by$^{13}$

\begin{equation}
\xi=\frac{2 \pi e^2}{\omega \hbar^2 c \sqrt{\kappa} }
\sum_{n n',\sigma \sigma'}
\int_0^{\infty} k dk |M_{n \sigma, n' \sigma'}(k) |^2 \delta(
\varepsilon_{n \sigma}(k) - \varepsilon_{n' \sigma'}(k) + \hbar \omega),
\end{equation}
where $\omega$ is the frequency of the infrared radiation,
$\kappa$ the dielectric permittivity, and $M_{n \sigma,
n' \sigma'}(k)$ the matrix elements for transitions between the
occupied subband states of the valence band and the empty subband states
of the conduction band ($n$ and $n'$ are the subband numbers):

\begin{eqnarray}
|M_{n \sigma, n' \sigma'}(k) |^2&=& \delta_{\sigma \sigma'} \left[
({\bf e} \cdot {\bf z})^2 P_{||}^2 \Phi_{n\sigma,n'\sigma'}^{(-)~2}(k)
+ (1/2) [{\bf e} \times {\bf z}]^2 P_{\bot}^2 \Phi_{n \sigma, n'
\sigma'}^{(+)~2}(k) \right] \nonumber \\
&&+(1/2) (1-\delta_{\sigma \sigma'}) [{\bf e} \times {\bf z}]^2
P_{\bot}^2 \Phi_{n \sigma, n' \sigma'}^{(+)~2}(k).
\end{eqnarray}
Here ${\bf e}$ is the unit vector of polarization of the infrared
radiation, ${\bf z}$ is the unit vector in the direction perpendicular
to the 2D plane, and
\begin{equation}
\Phi_{n\sigma,n'\sigma'}^{(\pm)}(k)
=\int_{-d/2}^{d/2} dz \left[ \varphi_{n
\sigma k}(z) \chi_{n' \sigma' k}(z) \pm  \chi_{n \sigma k}(z)
\varphi_{n' \sigma' k}(z) \right]
\end{equation}
are the overlap integrals. By taking the limits from $-d/2$ to
$d/2$, we neglect the barrier penetration of the wave functions,
i.e., we assume that the main part of the electron density remains
inside the well. The wave functions, therefore, should be normalized
according to
\begin{equation}
\int_{-d/2}^{d/2} dz \left[ \varphi_{n \sigma k}^2(z)+
\chi_{n \sigma k}^2(z) \right] = 1.
\end{equation}

Below we express the polarization-dependent absorption coeffecient as $\xi=
\xi_{\bot} [{\bf e} \times {\bf z}]^2 + \xi_{||} ({\bf e} \cdot {\bf z})$ and
calculate $\xi_{\bot}$ and $\xi_{||}$. In the case of normally incident
radiation one has $\xi= \xi_{\bot}$, regardless of the polarization. Only
two 2D levels, $n'=c$ and $n=v$, corresponding to the ground conduction- and
valence-band states, are involved in the calculation. Taking into account that
the energy levels of realistic physical systems are always broadened by
 disorder, we also introduce a finite broadening by replacing
the $\delta$-function in Eq. (15) by a Gaussian with a halfwidth of
1 meV. In this way we also avoid non-physical divergences of the optical
absorption coefficient.

Figure 6 shows the frequency dependence of the absorption coefficients
$\xi_{\bot}$ and $\xi_{||}$ for the quantum well described in Fig. 5.
Due to the small ratio of $(P_{||}/P_{\bot})^2$, it appears that $\xi_{||}$
is approximately ten times smaller than $\xi_{\bot}$. In zero field both
$\xi_{\bot}$ and $\xi_{||}$ show a step-like behavior at the fundamental
absorption edge$^{10}$, which is typical for 2D systems. In non-zero field,
when the densities of states for both valence and conduction subbands have
inverse-square-root divergences near the subband extrema, the absorption
shows sharp maxima near the threshold. This kind of behavior is typical
for 1D systems. Even after the introduction of a 1 meV broadening, the
absorption near the edge still remains considerably larger than the
zero-field absorption. It is important to note a qualitative difference
between the frequency dependences of $\xi_{\bot}$ and $\xi_{||}$ caused
by the fact that spin-flip transitions contribute to the first one
but   not to the second one. That is why $\xi_{\bot}$ shows jumps at
the frequencies corresponding to transitions with $k=0$, while $\xi_{||}$,
roughly following the density of states, shows bump-like features at
these frequencies.

Although the calculation according to Eq. (15) is done
under the assumption that the Fermi level remains in the gap, it is not
difficult to extend the formalism to the cases when the Fermi level remains
inside the valence or conduction band. A result of the calculation of
$\xi_{\bot}$ and $\xi_{||}$ at $F=10^5$ V/cm for a $n$-type structure with
$n=10^{11}$ cm$^{-2}$ is also shown in Fig. 6. Due to the Pauli exclusion
principle the maximum of the absorption at the threshold is suppressed.
Moreover, due to a considerable asymmetry of the spectra with respect to
their extrema points, cf. Fig 5. (a), there are actually two thresholds for
direct optical transitions in a $n$-type quantum well. The first one,
corresponding  approximately to $\hbar \omega = 11$ meV, signals a
transition with $k$ above the extremum point, while the second one,
corresponding approximately to $\hbar \omega = 17$ meV, signals a
transition with $k$ below the extremum point. For this reason
$\xi_{||}$ has two steps. On the other hand, the second step is not visible
in the frequency dependence of $\xi_{\bot}$. This is not
surprising because the overlap integrals (17) for $\xi_{||}$ and $\xi_{\bot}$
are different and it appears that transitions with $k$ below the extremum
point have a minor contribution to $\xi_{\bot}$ near the threshold.

Figure 7 shows the field dependence of the absorption of the same quantum
well at $\hbar \omega = 12$ meV. The absorption near the edge is very
sensitive to the electric field. This fact is important in connection
with possible applications of the biased narrow-gap quantum wells to
optical modulators. With the importance and high sensitivity of
differential spectroscopic measurements in mind, we also plot the
derivatives $d\xi_{\bot} /d F$ and $d\xi_{||} /d F$ in this figure.

We emphasize that the optical absorption coefficients calculated above
correspond to a single valley. Since in (111) grown
lead salt quantum wells there are two equivalent valleys described
by Eq. (1), both $\xi_{\bot}$ and $\xi_{||}$ should be multiplied
by a factor of 2.

\section{Conclusions}

We investigated the effect of a strong electric field on the energy spectrum
and the infrared optical properties of narrow-gap quantum wells described
by a two-band Dirac-like Hamiltonian. The corresponding eigenvalue problem
was solved numerically. Apart from an examination of an electric-field
control of the energy gap between the 2D states due to the Stark effect,
we established that a strong electric field transforms any narrow-gap
quantum well to a nearly gapless 2D system  with an electron energy
spectrum near the subband extrema   described by linear dispersion
laws. The physical reason for this behavior is described in terms of
an anticrossing of the conduction- and valence-subband spectra and
of the localization of the electron states near the well boundaries.
In general, the energy spectrum is dramatically modified by the
electric field, and the density of states, in addition to the
anticipated inverse-square-root divergency at the edges, has
similar divergences above the edges. The behavior of the density
of states manifests itself in the interband optical absorption.

In connection with the interest in far-infrared optics, we emphasize
the most important properties of the lead salt quantum wells caused by
an application of a strong electric field: (i) a possibility to control
the gap in a wide range including the far-infrared region, (ii) an
enchancement of the optical absorption near the fundamental edge,
and (iii) a high sensitivity of the absorption near the edge to the
electric field. Due to the Pauli exclusion principle, the enchancement
of the optical absorption would be suppressed in $n$ and $p$-type
quantum wells, where the Fermi level remains inside the conduction
or valence subbands.
The position of the Fermi level is not determined only
by the properties of the quantum well itself. Other factors, such as the
geometry
of the entire heterostructure and the remote doping profiles, are essential
as well. Therefore, by proper choice of parameters, one can achieve the
favorable conditions when the Fermi level is in the gap. However, the
question of a possible electric-field induced shift of the Fermi level
outside the gap still remains open.

Below we comment briefly on the approximations made in this paper.
In the description of the eigenvalue problem for narrow-gap lead salts
we restricted ourselves to the simple two-band Hamiltonian, neglecting
higher-band contributions, which in principle could be accounted for
with a set of Dimmock parameters$^5$. We neglected strain effects,
which introduce additional contributions$^6$ to the Hamiltonian.
We did not account for the difference between the velocities
$P_{||}$ and $P_{\bot}$ in the wells and barriers. Such a difference
would modify the boundary conditions, but if it is sufficiently small
it can be neglected. We considered homogeneous electric fields only
and neglected screening effects which may appear due to a redistribution
of the electron density in the wells. However, due to the high static
dielectric permittivity of the PbSnTe alloys, these effects should be
considerably less important than in GaAs-based structures.
Due to this high  dielectric permittivity we were also able to neglect the
Coulomb effects which otherwise would soften the sharp maxima of the
(pair-) density of states.
Further, we
considered (111)-grown heterostructures only. This choice is justified
because the lead salt heterostructures are typically grown on (111)
cleaved BaF$_2$ substrates$^{14}$. Finally, we neglected the broadening
of the subband levels due to escape of electrons from the wells by
field-assisted tunneling. Because of the small band offsets in
PbTe/SnTe systems, this effect is important and it essentially
restricts the range of the electric fields which can be applied
to the heterostructure. Nevertheless,
the use of PbSnTe quantum wells,
directly grown on BaF$_2$ or GaAs$^{15}$ substrates, would help to
extend this range to higher fields.

In conclusion, we have described profound modifications of the electron
energy spectra and density of states due to the anticrossing of 2D subbands
under strong electric fields.
The results were obtained by a direct numerical solution of Eq. (1) and,
to our knowledge have not been published before although the analytical solution
of Eq. (1), given in the Appendix, is known.
Similar modifications are possible not
only for lead salt quantum wells but also for systems
based  on other narrow-gap materials, for example, CdTe/HgTe and
InAs/GaSb heterostructures. Apart from their influence on the far-infrared
optical properties studied in this paper, the modifications of the spectra
must have a substantial effect on transport and equilibrium properties
of the narrow-gap low-dimensional electron systems.

\acknowledgements

This work was supported by le Minist\'{e}re de l' Education du
Qu\'{e}bec and by the Canadian NSERC Grant No. OGP0121756.

{\center {\bf APPENDIX\\}}

The general solution of Eq. (1) is expressed through the parabolic
cylinder functions$^{16}$ $D_j(x)$. With the definitions $\varphi
= (\tilde{ \varphi}+ i \tilde{ \chi})/\sqrt{2}$ and $\chi =
(\tilde{ \chi}+ i \tilde{ \varphi})/\sqrt{2}$ we have
$$
\tilde{ \varphi} = C_1 D_{i \beta -1} (\sqrt{i} s)
+ C_2 D_{i \beta -1} (-\sqrt{i} s),
\eqno (A1)
$$
$$
\tilde{ \chi} = (-\sqrt{i} \mu/\beta) \left[ C_1 D_{i \beta}
(\sqrt{i} s) - C_2 D_{i \beta} (-\sqrt{i} s) \right]
\eqno (A2)
$$
where $s=\sqrt{2 F/P_{||}}(z-\varepsilon/F)$, $\sqrt{i}=(1+i)
/\sqrt{2}$,
$
\beta=\left(\Delta^2/4 + P_{\bot}^2 k^2\right)/2 P_{||} F$, and
$\mu=\left(\sqrt{\Delta^2/4 + P_{\bot}^2 k^2}\right) / \left(- \Delta/2 +
i P_{\bot} \sigma k \right).
$
An application of the boundary conditions (2) to Eqs. (A1) and
(A2) allows to get an implicit equation describing the electron
spectrum. However, its analysis is more difficult than the direct
numerical solution of Eqs. (1) and (2) which we followed in
this paper.

\clearpage

\begin{figure}
\caption{Schematic band diagram of a biased narrow-gap quantum well. The
thick broken arrow shows an optical transition between
Stark-shifted 2D subbands.}
\label{fig.1}

\
\caption{Electron spectrum (a) and density of states per one valley (b)
for a narrow-gap quantum well with boundary conditions $\alpha_r=-\alpha_l=1$,
at $d=12$ nm and $\Delta=0$. Each graph is marked with a value of the
electric field $F$ in units of 10$^5$ V/cm.}
\label{fig.2}

\
\caption{The same as in Fig.2 at $\Delta=0.05$ eV.}
\label{fig.3}

\
\caption{Normalized wave functions $(\varphi, \chi)_{v,c}$ of ground
valence- and conduction-band states for a biased narrow-gap quantum well
at $\Delta=0$, $d=12$ nm, $F=2 \times 10^5$ V/cm, $\sigma=-1$ and
$k \simeq$ 0.175 nm$^{-1}$.
}
\label{fig.4}

\
\caption{Electron spectrum (a) and density of states per one valley (b)
of a 12 nm wide narrow-gap quantum well calculated with realistic boundary
conditions corresponding to SnTe/gapless PbSnTe interfaces. Each graph
is marked with a value of $F$ in units of 10$^5$ eV/cm.
}
\label{fig.5}

\
\caption{Frequency dependence of the absorption coeffecients $\xi_{\bot}$
(solid curve) and $\xi_{||}$ (magnified tenfold, dashed curve) for the quantum
well described in Fig. 5. Each graph is marked with a value of $F$ in units of
10$^5$ V/cm. A 1 meV broadening has been assumed.
The thin lines illustrate the Pauli exclusion effect on the absorption for
a $n$-type structure with $n=10^{11}$ cm$^{-2}$ at $F=10^5$ V/cm.
}
\label{fig.6}

\
\caption{Electric-field dependence of the far-infrared absorption coefficients
$\xi_{\bot}$ (solid curve) and $\xi_{||}$ (magnified tenfold, dashed curve)
for the
quantum well described in Fig. 5 at $\hbar \omega = 12$ meV. The first
derivatives of these functions (scaled and shifted for clarity) are shown
by thin lines.
}
\label{fig.7}

\end{figure}

\clearpage
\begin{references}
\bibitem[\dagger]{oleg}  E-mail: raichev@zinovi@lab2.kiev.ua
\bibitem[\diamond]{takis}  E-mail: takis@boltzmann.concordia.ca

\bibitem{1}Y. Huang,  J. Wang,  and C. Lien, J. Appl. Phys. {\bf 77}, 11 (1995).

\bibitem{2}S. M. Cho and H. H. Lee, J. Appl. Phys. {\bf 73}, 1918 (1993).

\bibitem{3}Z. Wang and J. K. Furdyna, J. Appl. Phys. {\bf 64}, 5248 (1988).

\bibitem{4}S. Hong, J. P. Loehr, J. E. Oh, P. K. Bhattachaya, and J. Singh,
Appl. Phys. Lett. {\bf 55}, 888 (1989).

\bibitem{5}G. Nimtz and B. Schlicht. Narrow-Gap Lead Salts, in {\em
Narrow-Gap Semiconductors}. Schpringer-Verlag, 1983.

\bibitem{6}M. Kriechbaum, K. E. Ambrosch, E. J. Fanter, H. Clemens, and G.Bauer
Phys. Rev. B {\bf 30}, 3394 (1984).

\bibitem{7}M. Kriechbaum, P. Kocevar, H. Pasher, and G. Bauer, IEEE J.
Quantum Electronics, {\bf 24}, 1727 (1988).

\bibitem{8}F. F. Sizov, V. V. Tetyorkin, and J. V. Gumenjuk-Sichevskaya,
Superlatt. Microstruct. {\bf 9}, 483 (1991).

\bibitem{9}V. K. Dugaev and P. P. Petrov. Fiz. Tekh. Poluprov. {\bf 23},
488 (1989) [Semiconductors, {\bf 23}, 305 (1989)];
Phys. Stat. Sol. (b) {\bf 184}, 347 (1994).

\bibitem{10}W. Okulski and M. Zaluzny, Thin Solid Films, {\bf 204}, 239 (1991).

\bibitem{11}F. T. Vasko and G. I. Steblovsky, Ukrainian Physical Journal
{\bf 34}, 576 (1989).

\bibitem{12}In theoretical papers the electron states in quantum
wells of narrow-gap lead salts are often described by an {\em
isotropic} Dirac Hamiltonian. In that case, the states in all eight
valleys are equivalent regardless of the growth direction. However,
the isotropic model is not satisfactory for an application to both
PbSnTe and PbSnSe materials.

\bibitem{13}F. T. Vasko and A. Kuznetsov, {\em Electronic States and Optical
Transitions in Semiconductor Heterostructures}. Springer, 1998.

\bibitem{14}H. Kinoshita and H. Fujiyasu, J. Appl. Phys. {\bf 51}, 5845 (1981).

\bibitem{15}K. T. Pollard, A. Erbil, and R. Sudharsanan, J. Appl. Phys. {\bf
71},
6136 (1992).

\bibitem{16}I. S. Gragshteyn and I. M. Ryzhik, {\em Tables of Integrals,
Series, and Products}. (Academic, New York, 1980).


\end{references}
\end{document}